\documentclass[pre,a4paper,showkeys,floatfix,superscriptaddress]{revtex4}

\usepackage{amssymb}
\usepackage{amsfonts}
 \usepackage{amsmath}
\usepackage{graphics}
\usepackage{palatino}
\usepackage{color}
\usepackage{epsfig}

\newcommand{\be}{\begin{equation}}
\newcommand{\ee}{\end{equation}}

\newcommand{\bea}{\begin{eqnarray}}
\newcommand{\eea}{\end{eqnarray}}

\begin{document}
\draft\preprint{Data incompleteness in TRNs}

\title{Topological effects of data incompleteness of gene regulatory networks.}

\author{J. Sanz}

\affiliation{Institute for Biocomputation and Physics of Complex Systems (BIFI), University of Zaragoza, Zaragoza 50009, Spain}
\affiliation{Department of Theoretical Physics, University of Zaragoza, Zaragoza 50009, Spain}

\author{E.Cozzo}

\affiliation{Institute for Biocomputation and Physics of Complex Systems (BIFI), University of Zaragoza, Zaragoza 50009, Spain}
\affiliation{Department of Condensed Matter Physics, University of Zaragoza, Zaragoza 50009, Spain}

\author{J. Borge-Holthoefer}

\affiliation{Institute for Biocomputation and Physics of Complex Systems (BIFI), University of Zaragoza, Zaragoza 50009, Spain}

\author{Y. Moreno}

\email{yamir.moreno@gmail.com}

\affiliation{Institute for Biocomputation and Physics of Complex Systems (BIFI), University of Zaragoza, Zaragoza 50009, Spain}
\affiliation{Department of Theoretical Physics, University of Zaragoza, Zaragoza 50009, Spain}

\begin{abstract}  
The topological analysis of biological networks has been a prolific topic in network science during the last decade. A persistent problem with this approach is the inherent uncertainty and noisy nature of the data. One of the cases in which this situation is more marked is that of transcriptional regulatory networks (TRNs) in bacteria. The datasets are incomplete because regulatory pathways associated to a relevant fraction of bacterial genes remain unknown. Furthermore, direction, strengths and signs of the links are sometimes unknown or simply overlooked. Finally, the experimental approaches to infer the regulations are highly heterogeneous, in a way that induces the appearance of systematic experimental-topological correlations. And yet, the quality of the available data increases constantly. In this work we capitalize on these advances to point out the influence of data (in)completeness and quality on some classical results on topological analysis of TRNs, specially regarding modularity at different levels. In doing so, we identify the most relevant factors affecting the validity of previous findings, highlighting important caveats to future prokaryotic TRNs topological analysis.
\end{abstract} 


\maketitle

\section{Introduction}

As it is commonly noticed in the literature, gene regulation is a complex process involving different phases and biochemical phenomenologies \cite{sorek2009prokaryotic,day1998post}. Among these mechanisms, transcriptional control constitutes one of the main resources the cell relies on to respond biochemically to environmental fluctuations and challenges. As a consequence, systematic characterization of TRNs has turned into a subject of high scientific interest \cite{sirbu2010}. Topological features of TRNs are customarily characterized at all scales using different metrics. At the large scale, genome-wide TRNs are signed and directed networks which present the following features: (i) regulatory proteins --origin of the regulatory interactions of the whole system-- represent a small fraction of the total number of nodes; (ii) out-going connectivity patterns are very heterogeneous --a small percentage of global regulators (hubs) send most of the links; and (iii) in-coming link distributions are quite compact: there is a characteristic scale that defines the typical number of regulations each protein receives \cite{babu2006evolutionary}.

Turning to the mesoscale, modularity appears also in TRNs as a key feature to understand the dynamical function of the system. In genome-wide TRNs, each regulator defines its own regulon as the set of nodes directly or indirectly regulated by it. Regulons are then subnetworks, that can be sometimes hierarchically organized; in other occasions, regulons partially overlap in non-trivial ways. Thus, the identification of groups of regulons --or parts of them-- interconnected through atypical, dense patterns is expected to store information about the biological role of the proteins within them \cite{bar2003computational}. The underlying idea is that community structure in biological networks might contribute to unveil functional modularity.

However, perhaps one of the most striking results on topological analysis of TRNs is related to small-scale (sets of 3 or 4 nodes) connectivity patterns, which present statistics anything but contingent \cite{milo2002}. Some of these patterns (or {\em motifs}) have been found to appear much more frequently than expected by random, while others, instead, are underrepresented in real networks. These statistical profiles, measured on different systems, allow the emergence of {\em network families}, each of which provide a general framework to understand the origin and the dynamical principles of the systems within them \cite{milo2004}. 

In addition to the aforementioned issues, the experimental challenges underlying the systemic characterization of the TRNs are far from being solved. The quantity and quality of available data on genome-wide transcriptional regulation are significant only for a small set of model organisms. Besides scarcity, the usual problem is related to the heterogeneous quality of the experimental evidences of the regulatory interactions, the building blocks of TRNs. Despite these problems, the amount of high-quality experimental information about transcriptional regulation at systemic level is growing each day, not only within the context of model prokaryotes. 

In this work, we analyze three of the best known prokaryotic TRNs, for which these data quality improvements are being more thoroughly incorporated to publicly available data sets. Two of them correspond to the model bacteria \emph{Escherichia coli} \cite{RegulonDB} and \emph{Bacillus subtilis} \cite{DBTBS}, while the third one corresponds to the pathogen \emph{Mycobacterium tuberculosis}, whose first network characterizations \cite{MtbRegList,Balazsi,sanz2011transcriptional} are more recent and incomplete due to the much higher difficulty associated to its wet-lab treatments and protocols. Specifically, the general question we set to answer here is whether robust and biologically relevant conclusions about TRNs can be reached given the current incompleteness of the data. Besides, we also show that some topological metrics do depend on the level of detail incorporated in TR maps, in particular, the structure of the mesoscale. Our findings show that extreme care should be taken when strong claims are made based on partial data. This is the case of TRNs superfamilies, which we argue are indeed grouped into a single class.  

\section{Community detection and link attributes}
The identification of modules in complex networks has attracted much attention of the scientific community in the last years. Algorithms and heuristics to optimize modularity ($Q$) \cite{newman04b} have appeared ever faster and more efficient \cite{fortunato2010community}, and generalizations to directed, weighted and signed networks are already available in the literature \cite{arenas08NJP,gomez09}. All these efforts have led to a considerable success regarding the quality of detected community structure in networks, and thus a more complete topological knowledge at this level has been attained. Behind this interest underlies the intuition that the relation between network structure and dynamics is strongly mediated by the mesoscale, and that community structure plays a central role in network formation and functioning. And yet, with few exceptions, link attributes are seldom taken into account. 

In this section we intend to underline that interaction direction and sign critically shape the detected community structure of a network. This is ever more dramatic in the case of TRNs, where a sharp distinction must be made between regulators (which mostly emit links) and the rest of the network, which mainly receive them. Also it is peculiar (though not exclusive) of these systems to allow for positive (activating) and negative (inhibitory) relationships. In practice, directions and signs are not always available in the datasets. Regarding directionality, we analyze a system --the TRN of {\em M. tuberculosis} \cite{sanz2011transcriptional}-- for which that is not an actual problem, as regulatory proteins are well identified, i.e. their function as link sources is known. Nevertheless, there are many cases of organisms whose regulatory pathways have not been explicitly identified, and in those cases the real topology is usually replaced by a co-expression network, which acts as an undirected proxy for the true underlying regulatory structure. Unavailability of interaction signs is, on the other hand, a more persistent problem: there exist many experimental approaches to infer a transcriptional regulation that do not inform about the sign of the interaction. Furthermore, there are interaction signs which depend on environmental conditions. Therefore, given the unavoidable incompleteness of the data, we explore whether link attributes determine the network modular structure, and to what extent.

To address the previous question, we perform a systematic comparison of the effects of preserving the original information (sign and direction) in modularity measures and community structure in TRNs. To this end, we will analyze the TRN of \emph{M. tuberculosis} \cite{sanz2011transcriptional}, for which we will consider three different topologies: one that preserves all available information (directed-signed, DS); an intermediate one (preserving directions, but not signs --directed-unsigned, DU); and a last one where all fine-grained information is ignored (undirected-unsigned, UU). From the output of this analysis, we provide a way to quantify how much biological information is lost when directions and/or signs are dropped out. Note that the three versions of the network have the same number of nodes $N$ and number of links $L$, the only differences being those regarding direction and/or the sign of the interactions. Interaction signs have been compiled from the experimental works enlisted in \cite{sanz2011transcriptional}, although signs were not reported there (see \cite{NotaWeb}).

The modularity expression used hereafter corresponds to its most general definition, i.e. the one that accounts for the existence of directions, weights, signed relations and self-loops, preserving the original information \cite{gomez09}:

\begin{equation}
Q = \frac{w^{+}}{w^{+}+w^{-}} Q^{+} - \frac{w^{-}}{w^{+}+w^{-}} Q^{-}
\label{eq:Q}
\end{equation}

This expression generalizes the concept of modularity, and simply computes the contribution to group formation of positive ($w^{+}$) and negative ($w^{-}$) interactions separately, $Q^{+}$ and $Q^{-}$ respectively; we refer to the original work for a full explanation of the expression. As for our current object of study, links in the network can only take values +1 or -1, and are originally defined as directed.

An intrinsic limitation of modularity maximization, as posed in Eq. \ref{eq:Q}, is that it provides a single snapshot of the modular structure of the network. However, several topological descriptions of the network coexist at different scales, which is, in general, a fingerprint of complex systems, and particularly relevant in biological ones \cite{Spirin}. A method to overcome this fundamental drawback of typical modularity optimization was put forth in \cite{arenas08NJP}. A parameter $r$ is introduced as a constant self-loop to each node, thus changing the total strength in a network and avoiding the inherent resolution limit of Newman's modularity $Q$ \cite{fortunato07resolution}. The shift only affects the property of each node individually and in the same way for all of them. Thus, the original adjacency matrix $\mathbf{A}$ is changed as a function of $r$: $\mathbf{A_{r}} = \mathbf{A} + \mathbf{Ir}$. The interesting property of the rescaled topology is that its characteristic scale in terms of modularity has changed. Then the topological structure revealed by optimizing the modularity for $\mathbf{A_{r}}$ is that of large groups for small values of $r$, and smaller groups for large values of $r$, all of which are strictly embedded in the original topology.

\begin{figure*}
  \begin{center}
  \includegraphics[width=0.5\columnwidth]{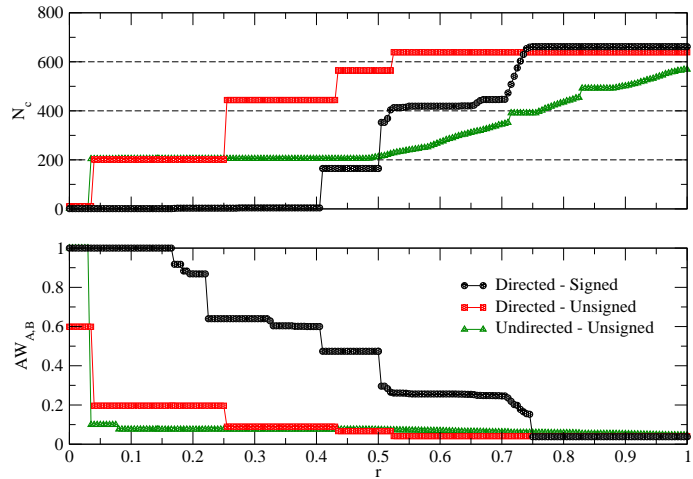}
  \end{center}
\caption{Top: number of detected modules $N_{c}$ as a function of the normalized rescaling parameter $r$. The mesoscale has been screened only for a range of $r$ yielding an interpretable amount of modules. DU and DS topologies show more than one persistent mesoscopic plateau, whereas the UU topology only has a single plateau made up of around $N_{c} \approx 200$ communities. Beyond $r=0.5$, no other stable plateau can be found for this topology. Bottom: the detected community structures for the three versions of the TRN of \emph{M. tuberculosis} are compared to the functional partitions in Tuberculist \cite{Tuberculist}. DS is the only network that shows significant values of similarity, in terms of the Asymmetric Wallace Index, against the functional partition for a large range of $r$ values. (Online version in colour.)}
\label{fig1}
\end{figure*}

Fig. \ref{fig1} (top) represents the number of modules $N_{c}$ that a combination of $Q$-maximization heuristics \cite{radatools} has detected for the three versions of the TRN of \emph{M. tuberculosis}. Each topology has been scrutinized at different scales, screening the parameter $r$ for 200 possible values, in a range such that it yielded an interpretable amount of modules. This range changes for different topologies, thus $r$ is normalized in the plot to allow for comparison. On visual inspection it is apparent that the three topologies present plateaus, where different $r$ values yield similar partitions in terms of $N_{c}$. This indicates that certain topological scales are robust and persistent, which might be a clue to identify functionally relevant groups of nodes \cite{arenas08NJP}. Notably, the UU topology presents a single plateau at $N_{c} = 205$ and then fails to stabilize for larger $r$'s. On the contrary, DU and DS, which retain more information, yield stable partitions at many levels. Although for different $r$ values, these topologies exhibit almost the same behavior regarding plateaus and the number of communities $N_{c}$ these plateaus present. At this point, one can say that the mesoscale analysis for DU and DS networks allows a richer interpretation in terms of the grouping of nodes, but there is no way to confirm if these are more or less biologically sound, than, for example, the UU topology.

To address this last question, we asked whether the partitions inferred by our method group genes with similar biological functions. The reason underlying this possibility is that genes within a topological community are connected among them by more regulations than average. This fact should imply that they tend to transcript together, as a response to common stimuli and eventually, to perform closely related functions. To do this, we compared the identified communities to the functional classification provided in the Tuberculist database \cite{Tuberculist}. There are many metrics and indices to compare two clusterings \cite{kuncheva04,rand,hubert,fowlkes,meila}. However, we need to rely on an index that does not severely punish different resolution scales: our reference partition categorizes genes in only $N_{c}^{F}=7$ groups, which yields a coarse-grained functional classification of the genome of \emph{M. tuberculosis} \cite{Notafunc}. Any partition with significantly more modules will show low resemblance to the functional one if the index is biased toward literally similar partitions. Thus, we present our results using the Asymmetric Wallace Index $AW$, which shows the inclusion of a partition into the other. The Asymmetric Wallace Index \cite{wallace83} is the probability that a pair of elements in one cluster of partition $A$ is also in the same cluster of partition $B$. Let be a clustering $A$ with $c_{A}$ communities and a clustering $B$ with $c_{B}$ communities, and let us define the confusion matrix $M$ whose rows correspond to the communities of the first clustering ($A$) and columns correspond to the communities of the second clustering ($B$). Let the elements of the confusion matrix, $M_{\alpha\beta}$, represent the number of common nodes between community $\alpha$ of the clustering $A$ and community $\beta$ of the clustering $B$; the partial sums being $M_{\alpha \cdot} = \sum_{\beta} M_{\alpha\beta}$ and $M_{\cdot\beta} = \sum_{\alpha} M_{\alpha\beta}$. Then, $AW_{A,B}$ (how much partition $A$ is embedded in $B$) is defined as follows:

\begin{equation}
AW_{A,B} = \frac{\displaystyle\sum_{\alpha=1}^{c_{A}} \displaystyle\sum_{\beta=1}^{c_{B}} M_{\alpha\beta} (M_{\alpha\beta} -1)}{\displaystyle\sum_{\alpha=1}^{c_{A}} M_{\alpha\cdot} -1}.
\end{equation}
The Asymmetric Wallace index can also be defined the other way around ($AW_{B,A}$), but in this case this is not considered, because detected partitions are systematically more divisive than the functional one, i.e. we are interested in seeing how detected partitions are embedded in the functional one.

Fig. \ref{fig1} (bottom) shows the results for the proposed scheme. Initial results (early $r$) for the UU and DS networks are artificially high, because $N_{c} < N_{c}^{F}$. Besides this, the plot indicates that only the partitions obtained from the DS topology are significantly similar to the functional one. In fact, beyond the initial stages of the resolution levels, both DU and UU's community structures are far from being embedded in the functional categorization. Quite surprisingly, resolution levels with similar $N_{c}$ do not entail similar $AW_{A,B}$ values. For instance, the three topologies show at some point a plateau with $N_{c} \approx 200$. But $AW_{UU,F} \approx 0.1$, $AW_{DU,F} \approx 0.2$ and finally $AW_{DS,F} \approx 0.5$.

These results suggest that the more complete knowledge about link attributes, the richer representation of the mesoscale, in which different levels of topological coarse-graining can be well identified, with possible bio-dynamical implications that need to be explored.

\section{Motifs significance robustness versus network growth}
Exhaustive search of topologically common footprints and systematic differences between different real systems constitutes an important topic in network theory since its very beginning \cite{Costa}. Along these lines, the classification of networks in families bring light into the evolutionary principles that ultimately yield to the complex topologies that real, evolving systems like TRNs show today \cite{babu2006evolutionary}. In this sense, the work by Alon and coworkers \cite{milo2004} constitutes a milestone.

In their work, the statistical significance of 3-nodes motifs --triads-- was analyzed. The number of appearances of each of the thirteen possible directed structures in real systems was compared to those observed in a null model. The null ensemble was constructed by randomly rewiring the links of the original networks, preserving the number of single links and mutual interactions \cite{Nota}. The statistical significance of each motif $h$ is then defined as the Z-score of its number of appearances when compared to the results found in the null ensemble:

\begin{equation}
Z_{\text{score}_{\text{h}}}= \frac{n_{h}-\langle n_{rand,h}\rangle}{\sigma_{rand,h}}
\label{eq1}
\end{equation}

Therefore, computing the $Z_{\text{score}}$ for all possible triads in a network yields a 13-dimensional vector that, when normalized, represents the so-called triad significance profile (TSP). From the analysis of different systems' profiles, four superfamilies were identified with common TSPs: two families of non-biological networks --semantic adjacency words maps and social systems-- and two families of biological, information processing networks.  

\begin{figure}
  \begin{center}
  \includegraphics[width=0.5\columnwidth]{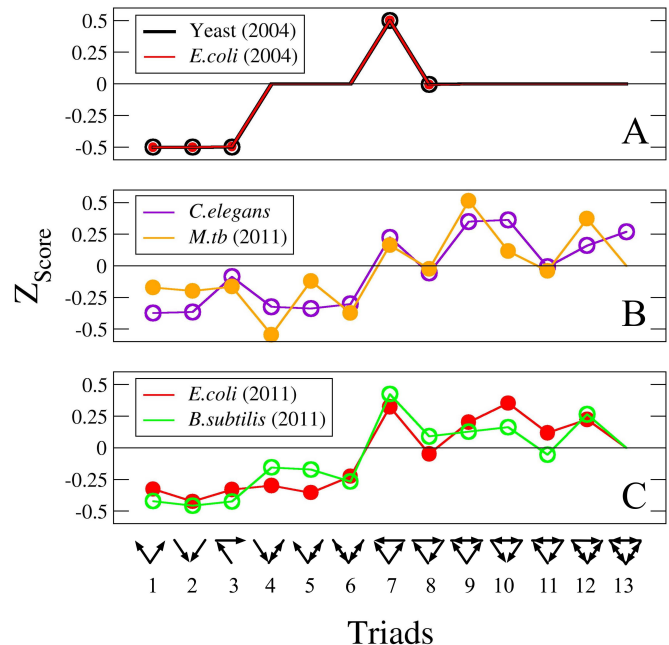}
  \end{center}
\caption{Triad significance profiles (TSPs) of bio-information processing networks. Panel A: \emph{E.coli} (2004). TRN of \emph{E.coli} as firstly published in \cite{WebAlon,Mangan} ($N=423$ operons, $L=519$ regulations plus $SL=59$ self-regulations). Yeast (2004) TRN of budding yeast \cite{milo2002} ($N=688$ genes, $L=1079$ regulations \cite{WebAlon,Notayeast}). Panel B: \emph{C.elegans} neural network \cite{Celegans} ($N=279$ neurons, $L=2990$ synapses).  \emph{M.tuberculosis} TRN \cite{sanz2011transcriptional} ($N=1624$ genes, \cite{notagenes} $L=3169$ regulations plus $SL=43$ self-regulations). Panel C: \emph{E.coli} (2011). Updated TRN of \emph{E.coli} based on RegulonDB, release 7.2, dated on May, 2011 ($N=1037$ operons, $L=2574$ regulations plus $SL=113$ self-regulations).  \emph{B.subtilis} (2011), updated TRN of \emph{B.subtilis}, based on DBTBS database \cite{DBTBS}, (accessed in October, 2011) ($N=814$ operons, $L=1294$ regulations plus $SL=80$ self-regulations).(Online version in colour.)}
\label{fig2}
\end{figure}

Regarding the two biological networks superfamilies originally identified, TRNs of three unicellular organisms were found to conform the first one: yeast, \emph{B.subtilis} and \emph{E.coli}. In Fig. \ref{fig2}, panel A, we plot the TSPs that belong to two of the four datasets analyzed by the authors in their original work: yeast \cite{Costanzo} and \emph{E.coli} (available at the authors' web site \cite{WebAlon}). The second group contains developmental TRNs of eukaryotic cells belonging to pluricellular organisms, signal transduction maps and synaptic networks. In Fig. \ref{fig2}, panel B, the TSP of the synaptic wiring map of the nematode \emph{C.elegans} \cite{Celegans} is plotted, as an example of this second superfamily, evidencing the differences with respect to panel A. 

The biological interpretation of the emergence of the two superfamilies of TRNs --or more generally, bio-information processing networks-- proposed in \cite{milo2004} has to do with the typical response times developed by each group of systems. These times are similar to those of single interactions for the networks in the first group (rate-limited networks) but remarkably greater than characteristic interaction times for the systems within the second superfamily (unrate-limited networks). 

The recent addition to this scheme of the TRN of \emph{M.tuberculosis} poses an intriguing question. As it is visible to the naked eye in Fig. \ref{fig2} (panel B) its TSP, although belonging to an unicellular organism, has a greater correlation with the representative of the unrate-limited superfamily. The fact that \emph{M.tb.} has these developmental-like topological features at its TRN might be interpreted under a coherent biological picture \cite{sanz2011transcriptional}. The pathogen has an evolutive history tightly bound to its condition of a human intracellular obligate parasite, which could eventually have caused an adaptation of the bacterium to the rhythms and response dynamics of host cells. Indeed, certain stimuli, like hypoxia, yield anomalously slow shifts in \emph{Mycobacterium tuberculosis} gene expression patterns, which can take as much as 80 days until stabilization \cite{Balazsi}.

The third panel in Fig. \ref{fig2} invalidates the previous hypothesis, and presents the TSPs of the updated TRNs of two bacteria which were initially characterized as rate-limited according to their TSPs. Visible at a glance, the update of the datasets has shifted their TSPs from one superfamily to another, in a way that suggests that the division of the information processing networks into two groups was an effect of data incompleteness.

The key of the change observed in the TSPs stems from the small number of two nodes feedback loops that are observed in unicellular organisms TRNs \cite{Nota2}. When feedbacks are absolutely absent from the system under study, as the randomizing algorithm preserves the number of them, feedback loops will also be absent in the null ensemble. This situation makes the Z-scores associated to triads 4, 5, 6, 9, 10, 11, 12 and 13 undefined, as in Fig. \ref{fig2}, panel A. As time goes by, such cases have become obsolete: new links have been discovered and added to the growing datasets, and some of them generate feedback loops, which are now present in the triads listed before. In the three updated systems studied, we have found as many as 12 feedback loops in \emph{E.coli} TRN, 9 in \emph{B.subtilis} and 6 in \emph{M.tuberculosis}. The result, after the incorporation of these new feedbacks, suppose that the division between two superfamilies of biological information processing networks according to their TSPs disappears, affecting the biological interpretation about the eventual relationship between time responses and motifs statistics. 

\section{Systematic correlations between topology and experimental evidence}

Experimental techniques used in transcriptional regulation inference are numerous and often subtle \cite{Banerjee}. However, usual approaches can be grouped within two main categories. The first approach is based on the explicit detection of the physical protein-DNA interaction between regulators and promoters of target genes. This presents the advantage that only direct operations of regulators on targets can be observed. However, the existence of a protein-DNA interaction under certain in-vitro conditions does not guarantee that it is physiologically relevant in terms of target expression levels. 

\begin{table}[tdp]
\caption{Number of links reported based upon binding assays (BA) or target expression levels comparison (TELC). Well characterized links (WCLs) are those characterized at least by one methodology of each group (for an exhaustive list of the experimental methods in each group see \cite{NotaWeb}). Poorly characterized links (PCLs) are reported under methodologies that can not be considered within neither of the two main groups (too generic methods, orthologies based deductions, absence of experimental support etc.). Whilst \emph{B.subtilis} and \emph{E.coli} are relatively well characterized systems, in order to get enough statistics for the \emph{M.tuberculosis} case, we consider identification of consensus sequences as binding proofs.}
\begin{center}
\begin{tabular}{c|c|c|c}
&  \emph{E.coli} & \emph{B.subtilis} & \emph{M.tuberculosis}\\\hline
BA & 914 & 499 & 726 \\\hline
TELC & 1272 & 856 & 1290 \\\hline
WCLs & 656 & 323 & 191 \\\hline
PCLs & 1044 & 262 & 1344 \\\hline
\end{tabular}
\end{center}
\label{tabla3}
\end{table}%

The alternative approach is essentially based on the generation of mutant strains in which the functionality and/or the expression levels of a certain binding factor are significantly altered with respect to those of the wild type. Then, expression levels of genes which are potentially regulated by the binding factor under study are registered and compared between wild type and mutant strains. In this way, if these different levels of regulator activity yield significantly different target expression measures, one might assume that the regulator is actually acting on the target. 

The main advantage of the latter approach is that the sign and strength of the interaction can be determined. However, the analysis cannot distinguish direct regulatory interactions from indirect influences regulator-target mediated by secondary regulatory pathways. Nonetheless, as it can be seen in Table \ref{tabla3}, this second kind of methods is responsible for the characterization of an important fraction of the links in our systems. Therefore, a relevant question is whether or not the appearance of indirect, spurious links (as if they were real interactions) might suppose a systematic error responsible of topological bias at a global level.

\begin{figure}
  \begin{center}
  \includegraphics[width=0.5\columnwidth]{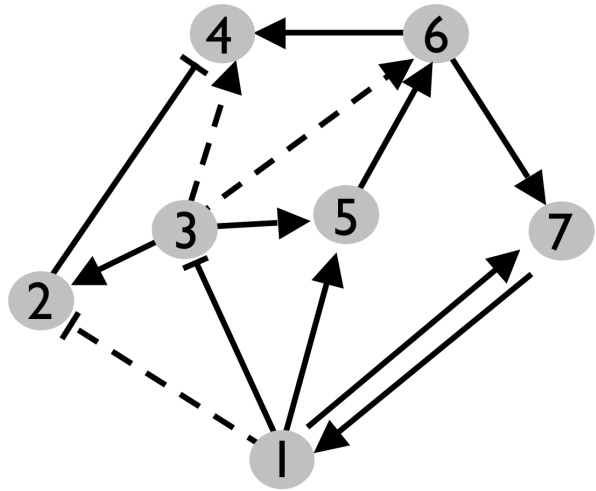}
  \end{center}
\caption{Schematic representation of suspicious links. In the figure, arrows represent activations and right angles inhibitions. Dotted lines correspond to the links that are topologically suspicious, i.e., those for which a secondary regulatory pathway mediated by a third gene is registered and whose sign is coherent. For example, the link connecting nodes 1 and 2 is considered suspicious because of the existence of the pathway 1 to 3 and 3 to 2. The same happens for the link between nodes 3 and 4. In both cases, the condition that the product of the two links making the secondary pathway should coincide with that of the link being considered as suspicious is verified. In this sense, we have also included a case in which the latter condition does not hold: the edge linking node 1 to node 5 is not suspicious because although a two-nodes cascade connecting the same nodes exists, (i.e. 1 to 3 and 3 to 5) it is not sign coherent.}
\label{fig3}
\end{figure}

These hypothetical spurious interactions should appear connecting nodes for which a secondary regulation pathway exists, and its sign should be the same of that secondary route (see Fig. \ref{fig3}). So, in our networks, we can identify those ``suspicious'' links (SLs) connecting nodes for which some secondary via has been already registered, and verify for sign coherence. We will restrict our analysis to those secondary pathways formed by a two-links cascade. The question is how we can know whether this subset of suspicious links presents a higher rate of spurious links than on average. Indeed, among the topologically suspicious links, only those that are characterized by at least one technique of each methodological category --henceforth referred to as well-characterized links (WCLs)--, can safely be considered as non-suspicious direct regulations.

\begin{table}[tdp]
\caption{\emph{Suspicious} links are that could be annotated as a result of the existence of a sign coherent secondary pathway (see Fig. \ref{fig3}). Background percent of WCLs are calculated in relation to the total number of signed links.The null hypothesis $H_{o}$ assumes that the proportion of WCLs within the SLs subset should obey a normal distribution centered at the background value.}
\begin{center}
\begin{tabular}{c|c|c|c}
&  \emph{E.coli} & \emph{B.subtilis} & \emph{M.tuberculosis}\\\hline
SLs & 732 & 231 & 263 \\\hline
SLs which are WCLs & 104 (14.2\%) & 43 (18.6\%) & 11 (4.2\%) \\\hline
Total, signed WCLs & 656 (25.5\%) & 323 (25.1\%) & 191 (7.4\%) \\\hline
$H_{o}$ $p$ value & $<10^{-10}$ & $0.024$ & $0.028$ \\\hline
\end{tabular}
\end{center}
\label{tabla4}
\end{table}%

Therefore, the idea is to compare the proportion of well characterized links within the subset of topologically suspicious links with the same proportion but for the whole network. In Table \ref{tabla4}, we annotate the proportion of these more reliable, well characterized links within the subset of suspicious interactions compared to the same fraction of WCLs in the whole signed networks. As it can be seen, suspicious links systematically present a slightly lower proportion of WCLs than the background, which could be associated to random fluctuations with respect to the average values with probabilities lower than 3\% in each of the systems, being remarkably lower in the case of the TRN of \emph{E.coli}.

This indicates that suspicious links constitute a topologically defined subset of interactions which is systematic and significantly less reliably characterized than on average in all the systems under study. This observation is in agreement with the hypothesis that insufficient experimental methods of transcriptional regulation inference can suppose the systematical observation of topologically-biased spurious links. The problem addressed here seems to critically affect the characterization of the activity of sigma factors. In fact, when we reconstruct the networks under study by considering only transcription factors as regulators and exclude sigma factors, the whole picture significantly changes. Indeed, the percent of suspicious links which are better characterized is even greater than the background, both for \emph{B.subtilis} ($45.0\%$ vs $42.2\%$) and for \emph{E.coli} ($46.0\%$ vs $43.1\%$). For the case of \emph{M.tuberculosis}, the analysis can be hardly conclusive due to the loss of statistics after sigma factors removal (no well characterized link is located within the set of suspicious interactions, now, less than 100 in the whole signed network). These findings, put together, suggest that characterization of sigma factor regulons is more sensitive to the aforementioned issues. 

\begin{table}[tdp]
\caption{Changes in the $Z_{\text{scores}}$ of the most affected motifs -cascades and feed-forward loops (FFLs)- due to systematic mischaracterization of suspicious links. Note the almost perfect symmetry of the values, which is not affected by the correction procedure.}
\begin{center}
\begin{tabular}{c|c|c|c}
&  \emph{E.coli} & \emph{B.subtilis} & \emph{M.tb.}\\\hline
Cascade (original) & $-4.7\pm 0.2$ & $-6.9\pm 0.4$ & $-2.2\pm 0.1$ \\
Cascade (corrected) & $-2.9\pm 0.4$ & $-6.9\pm 0.8$ & $-1.1\pm 0.3$ \\\hline
FFL (original) & $4.7\pm 0.2$ & $6.9\pm 0.4$ & $2.2\pm 0.1$ \\
FFL (corrected) & $2.5\pm 0.7$ & $7.0\pm 0.8$ & $1.1\pm 0.3$ \\\hline
\end{tabular}
\end{center}
\label{tabla5}
\end{table}%

\begin{figure}
  \begin{center}
  \includegraphics[width=0.5\columnwidth]{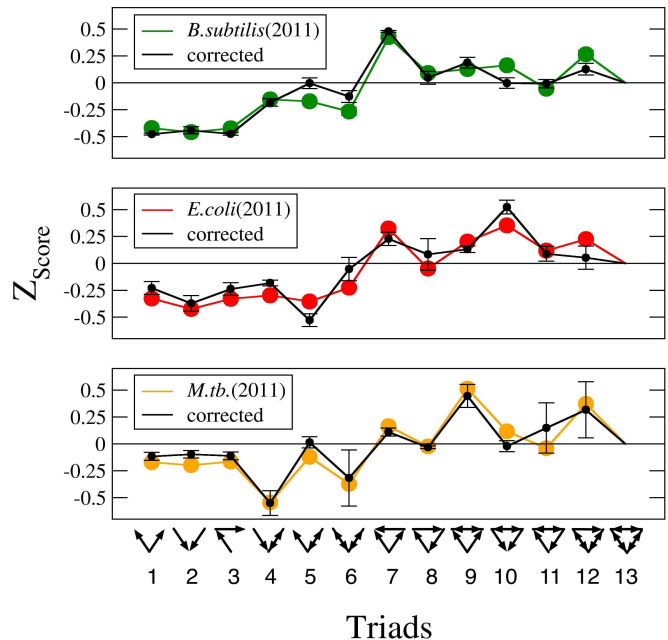}
  \end{center}
\caption{Changes in prokaryotic TSPs due to systematic experimental mischaracterization of links. Original systems present a lower proportion of WCLs than the background in the set of links that connect nodes for which a secondary coherent regulatory pathway has been registered. In corrected networks, the proportion of WCLs is paired to background levels by randomly deleting poorly characterized interactions. This, however, do not affect the TSPs significantly.(Online version in colour.)}
\label{fig4}
\end{figure}

Another issue of interest is whether this experimental bias is topologically relevant. More precisely, we question if this systematic error could quantitatively affect motif statistics in our systems. The key is that these spurious interactions could recurrently transform some motifs into others, and more precisely, focusing on most prominent motifs in number of appearances, this would suppose the systematic, spurious transformation of three-nodes cascades (triad 3) into coherent feedforward loops (triad 7). To test the robustness of the TSPs to the presence of spurious links, we delete in each network a fraction of partially characterized suspicious links up to the point in which the proportion of WCLs among them is comparable to the average background level. This suppose the removal of 324 suspicious links in the TRN of \emph{E.coli}, 59 in the TRN of \emph{B.subtilis} and 114 for the \emph{M. tuberculosis} case. The links to be deleted are randomly chosen within the set of partially characterized suspicious links. Finally, we recalculate the $Z_{\text{scores}}$ of all the motifs and compare TSPs with their original values. The results of this process are shown in Table \ref{tabla5}, where it can be seen that the statistical significance of cascades and feedforward loops are systematically affected. The interesting fact is that, after the correction, cascades are yet significantly underrepresented while feedforward loops as a whole (i.e. independently of the signs) continue to appear much more frequently than expected by random. Obviously, the $Z_{\text{score}}$ associated to the other triads also varies. But the striking point is that, after normalization, in all the systems the effect of the correction on the TSPs are very limited, as we can see in Fig. \ref{fig4}. So, the conclusion is that this kind of systematic error, although modifying the absolute values of motifs' $Z_{\text{scores}}$, does not affect their ratios that are recovered after normalizing the TSPs.

\section*{Conclusions}

As we have shown here, sources of unreliability can be of diverse nature: from the often unjustified lack of details in link attributes to the lack of key interactions, whose inclusion radically modify motifs' TSPs. As a matter of fact, our first finding convincingly shows that data incompleteness could exert a relevant influence on the topological characterization of the mesoscale in prokaryotic TRNs. More precisely, we have shown how a complete knowledge of link attributes (directions and signs) can yield richer mesoscale structures in TRNs. Secondly, we have also shown that a mere updating of the interactions that make up a TRN in which key regulatory interactions are incorporated, radically modifies previous results based on the analysis of motifs appearances. In fact, some of the previous conclusions do not hold anymore. We have observed that prokaryotic TRNs show motifs significance profiles very similar to those belonging to multicellular, developmental TRNs, signal transduction and neural systems. Finally, experimental mischaracterization of the links has also been studied, and yet, we have found that its influence on motifs statistics is reduced. These results suggest that the evolutionary interplay between topology and dynamics is more similar between regulatory systems of multicellular and unicellular organisms than expected.

Transcriptional Regulatory Networks have been increasingly studied during the last several years. Nowadays, however, their characterization can only be considered provisional, as they consist of incomplete annotations of often heterogeneous and unreliable experimental evidences, computational inferences and theoretical predictions. While working with still incomplete networks could be of valuable help to uncover unknown biochemical pathways, there are situations in which reliable conclusions cannot be obtained. Moreover, we don't even know when the latter is the case. Accuracy and robustness of the results thus require us to be able to assess what results are dependent on the noisy and uncertain nature of some annotated links. This is crucial if deep biological implications are to be claimed. 

\begin{acknowledgments}
We would like to thank Jos\'e Alberto Carrodeguas and Alejandra Nelo for helpful comments. This work has been partially supported by MICINN through Grants FIS2008-01240, FIS2009-13364-C02-01, and FIS2011-25167 and by Comunidad de Arag\'on (Spain) through a grant to FENOL group.
\end{acknowledgments}


\begin{thebibliography}{99}
\bibitem{Notafunc} For the comparison between topological partitions and functional classification only genes belonging to truly structural functions have been considered. So, ``conserved hypotheticals'' and ``unknown'' genes have been excluded, as well as genes with regulatory roles (``regulatory proteins'' and ``information pathways'' genes), which are expected to join transversally the TRN.
\end{thebibliography}
\end{document}